\documentclass[journal,draftclsnofoot,onecolumn,12pt]{IEEEtran}

\usepackage{graphicx}
\usepackage{subfigure}
\usepackage{multirow}
\usepackage{array}
\newcolumntype{P}[1]{>{\centering\arraybackslash}p{#1}}
\newcolumntype{M}[1]{>{\centering\arraybackslash}m{#1}}

\usepackage{amsthm,amssymb,amsmath,graphicx,multirow,color,amsfonts}%
\usepackage[update,prepend]{epstopdf}
\usepackage{multirow}
\usepackage[latin1]{inputenc}
\usepackage{tikz}
\usepackage{bbm} 
\usepackage{pdfpages}
\usepackage{multirow}
\usepackage{subfig}
\usepackage{comment}
 \usepackage{graphicx}
\usepackage{multicol}
\usepackage{cite}

\usepackage[justification=centering]{caption}
\usepackage{textcomp}
\usepackage{psfrag}
\usepackage{arydshln}
\usepackage{url}
\usepackage{soul}
\usepackage{graphicx,color}
\usepackage[nolist]{acronym}
\usepackage{algorithm,algorithmic} 

\usepackage{mathtools,lipsum}
\usepackage{cuted}
\usepackage{amsmath}
 
\usepackage{mathrsfs}

\usepackage[capitalise]{cleveref}
\Crefname{equation}{Eq.\!}{Eqs.\!}
\Crefname{figure}{Fig.\!}{Figs.\!}
\Crefname{tabular}{Tab.\!}{Tabs.\!}
\Crefname{section}{Section\!}{Sections.\!}

\def\nb0{{\mathbf{0}}}
\def\nb1{{\mathbf{1}}}

\newtheorem{lemma}{Lemma}

\newtheorem{definition}{Definition}

\newtheorem{theorem}{Theorem}

\newtheorem{corollary}{Corollary}

\newenvironment{sequation}{
\begin{equation}\small}{\end{equation}
}

\begin{document}
\graphicspath{{./Figures/}}
	\begin{acronym}

\acro{5G-NR}{5G New Radio}
\acro{3GPP}{3rd Generation Partnership Project}
\acro{ABS}{aerial base station}
\acro{AC}{address coding}
\acro{ACF}{autocorrelation function}
\acro{ACR}{autocorrelation receiver}
\acro{ADC}{analog-to-digital converter}
\acrodef{aic}[AIC]{Analog-to-Information Converter}     
\acro{AIC}[AIC]{Akaike information criterion}
\acro{aric}[ARIC]{asymmetric restricted isometry constant}
\acro{arip}[ARIP]{asymmetric restricted isometry property}

\acro{ARQ}{Automatic Repeat Request}
\acro{AUB}{asymptotic union bound}
\acrodef{awgn}[AWGN]{Additive White Gaussian Noise}     
\acro{AWGN}{additive white Gaussian noise}

\acro{APSK}[PSK]{asymmetric PSK} 

\acro{waric}[AWRICs]{asymmetric weak restricted isometry constants}
\acro{warip}[AWRIP]{asymmetric weak restricted isometry property}
\acro{BCH}{Bose, Chaudhuri, and Hocquenghem}        
\acro{BCHC}[BCHSC]{BCH based source coding}
\acro{BEP}{bit error probability}
\acro{BFC}{block fading channel}
\acro{BG}[BG]{Bernoulli-Gaussian}
\acro{BGG}{Bernoulli-Generalized Gaussian}
\acro{BPAM}{binary pulse amplitude modulation}
\acro{BPDN}{Basis Pursuit Denoising}
\acro{BPPM}{binary pulse position modulation}
\acro{BPSK}{Binary Phase Shift Keying}
\acro{BPZF}{bandpass zonal filter}
\acro{BSC}{binary symmetric channels}              
\acro{BU}[BU]{Bernoulli-uniform}
\acro{BER}{bit error rate}
\acro{BS}{base station}
\acro{BW}{BandWidth}
\acro{BLLL}{ binary log-linear learning }

\acro{CP}{Cyclic Prefix}
\acrodef{cdf}[CDF]{cumulative distribution function}   
\acro{CDF}{Cumulative Distribution Function}
\acrodef{c.d.f.}[CDF]{cumulative distribution function}
\acro{CCDF}{complementary cumulative distribution function}
\acrodef{ccdf}[CCDF]{complementary CDF}               
\acrodef{c.c.d.f.}[CCDF]{complementary cumulative distribution function}
\acro{CD}{cooperative diversity}

\acro{CDMA}{Code Division Multiple Access}
\acro{ch.f.}{characteristic function}
\acro{CIR}{channel impulse response}
\acro{cosamp}[CoSaMP]{compressive sampling matching pursuit}
\acro{CR}{cognitive radio}
\acro{cs}[CS]{compressed sensing}                   
\acrodef{cscapital}[CS]{Compressed sensing} 
\acrodef{CS}[CS]{compressed sensing}
\acro{CSI}{channel state information}
\acro{CCSDS}{consultative committee for space data systems}
\acro{CC}{convolutional coding}
\acro{Covid19}[COVID-19]{Coronavirus disease}

\acro{DAA}{detect and avoid}
\acro{DAB}{digital audio broadcasting}
\acro{DCT}{discrete cosine transform}
\acro{dft}[DFT]{discrete Fourier transform}
\acro{DR}{distortion-rate}
\acro{DS}{direct sequence}
\acro{DS-SS}{direct-sequence spread-spectrum}
\acro{DTR}{differential transmitted-reference}
\acro{DVB-H}{digital video broadcasting\,--\,handheld}
\acro{DVB-T}{digital video broadcasting\,--\,terrestrial}
\acro{DL}{DownLink}
\acro{DSSS}{Direct Sequence Spread Spectrum}
\acro{DFT-s-OFDM}{Discrete Fourier Transform-spread-Orthogonal Frequency Division Multiplexing}
\acro{DAS}{Distributed Antenna System}
\acro{DNA}{DeoxyriboNucleic Acid}

\acro{EC}{European Commission}
\acro{EED}[EED]{exact eigenvalues distribution}
\acro{EIRP}{Equivalent Isotropically Radiated Power}
\acro{ELP}{equivalent low-pass}
\acro{eMBB}{Enhanced Mobile Broadband}
\acro{EMF}{ElectroMagnetic Field}
\acro{EU}{European union}
\acro{EI}{Exposure Index}
\acro{eICIC}{enhanced Inter-Cell Interference Coordination}

\acro{FC}[FC]{fusion center}
\acro{FCC}{Federal Communications Commission}
\acro{FEC}{forward error correction}
\acro{FFT}{fast Fourier transform}
\acro{FH}{frequency-hopping}
\acro{FH-SS}{frequency-hopping spread-spectrum}
\acrodef{FS}{Frame synchronization}
\acro{FSsmall}[FS]{frame synchronization}  
\acro{FDMA}{Frequency Division Multiple Access}

\acro{GA}{Gaussian approximation}
\acro{GF}{Galois field }
\acro{GG}{Generalized-Gaussian}
\acro{GIC}[GIC]{generalized information criterion}
\acro{GLRT}{generalized likelihood ratio test}
\acro{GPS}{Global Positioning System}
\acro{GMSK}{Gaussian Minimum Shift Keying}
\acro{GSMA}{Global System for Mobile communications Association}
\acro{GS}{ground station}
\acro{GMG}{ Grid-connected MicroGeneration}

\acro{HAP}{high altitude platform}
\acro{HetNet}{Heterogeneous network}

\acro{IDR}{information distortion-rate}
\acro{IFFT}{inverse fast Fourier transform}
\acro{iht}[IHT]{iterative hard thresholding}
\acro{i.i.d.}{independent, identically distributed}
\acro{IoT}{Internet of Things}                      
\acro{IR}{impulse radio}
\acro{lric}[LRIC]{lower restricted isometry constant}
\acro{lrict}[LRICt]{lower restricted isometry constant threshold}
\acro{ISI}{intersymbol interference}
\acro{ITU}{International Telecommunication Union}
\acro{ICNIRP}{International Commission on Non-Ionizing Radiation Protection}
\acro{IEEE}{Institute of Electrical and Electronics Engineers}
\acro{ICES}{IEEE international committee on electromagnetic safety}
\acro{IEC}{International Electrotechnical Commission}
\acro{IARC}{International Agency on Research on Cancer}
\acro{IS-95}{Interim Standard 95}

\acro{KPI}{Key Performance Indicator}

\acro{LEO}{low earth orbit}
\acro{LF}{likelihood function}
\acro{LLF}{log-likelihood function}
\acro{LLR}{log-likelihood ratio}
\acro{LLRT}{log-likelihood ratio test}
\acro{LoS}{Line-of-Sight}
\acro{LRT}{likelihood ratio test}
\acro{wlric}[LWRIC]{lower weak restricted isometry constant}
\acro{wlrict}[LWRICt]{LWRIC threshold}
\acro{LPWAN}{Low Power Wide Area Network}
\acro{LoRaWAN}{Low power long Range Wide Area Network}
\acro{NLoS}{Non-Line-of-Sight}
\acro{LiFi}[Li-Fi]{light-fidelity}
 \acro{LED}{light emitting diode}
 \acro{LABS}{LoS transmission with each ABS}
 \acro{NLABS}{NLoS transmission with each ABS}

\acro{MB}{multiband}
\acro{MC}{macro cell}
\acro{MDS}{mixed distributed source}
\acro{MF}{matched filter}
\acro{m.g.f.}{moment generating function}
\acro{MI}{mutual information}
\acro{MIMO}{Multiple-Input Multiple-Output}
\acro{MISO}{multiple-input single-output}
\acrodef{maxs}[MJSO]{maximum joint support cardinality}                       
\acro{ML}[ML]{maximum likelihood}
\acro{MMSE}{minimum mean-square error}
\acro{MMV}{multiple measurement vectors}
\acrodef{MOS}{model order selection}
\acro{M-PSK}[${M}$-PSK]{$M$-ary phase shift keying}                       
\acro{M-APSK}[${M}$-PSK]{$M$-ary asymmetric PSK} 
\acro{MP}{ multi-period}
\acro{MINLP}{mixed integer non-linear programming}

\acro{M-QAM}[$M$-QAM]{$M$-ary quadrature amplitude modulation}
\acro{MRC}{maximal ratio combiner}                  
\acro{maxs}[MSO]{maximum sparsity order}                                      
\acro{M2M}{Machine-to-Machine}                                                
\acro{MUI}{multi-user interference}
\acro{mMTC}{massive Machine Type Communications}      
\acro{mm-Wave}{millimeter-wave}
\acro{MP}{mobile phone}
\acro{MPE}{maximum permissible exposure}
\acro{MAC}{media access control}
\acro{NB}{narrowband}
\acro{NBI}{narrowband interference}
\acro{NLA}{nonlinear sparse approximation}
\acro{NLOS}{Non-Line of Sight}
\acro{NTIA}{National Telecommunications and Information Administration}
\acro{NTP}{National Toxicology Program}
\acro{NHS}{National Health Service}

\acro{LOS}{Line of Sight}

\acro{OC}{optimum combining}                             
\acro{OC}{optimum combining}
\acro{ODE}{operational distortion-energy}
\acro{ODR}{operational distortion-rate}
\acro{OFDM}{Orthogonal Frequency-Division Multiplexing}
\acro{omp}[OMP]{orthogonal matching pursuit}
\acro{OSMP}[OSMP]{orthogonal subspace matching pursuit}
\acro{OQAM}{offset quadrature amplitude modulation}
\acro{OQPSK}{offset QPSK}
\acro{OFDMA}{Orthogonal Frequency-division Multiple Access}
\acro{OPEX}{Operating Expenditures}
\acro{OQPSK/PM}{OQPSK with phase modulation}

\acro{PAM}{pulse amplitude modulation}
\acro{PAR}{peak-to-average ratio}
\acrodef{pdf}[PDF]{probability density function}                      
\acro{PDF}{probability density function}
\acrodef{p.d.f.}[PDF]{probability distribution function}
\acro{PDP}{power dispersion profile}
\acro{PMF}{probability mass function}                             
\acrodef{p.m.f.}[PMF]{probability mass function}
\acro{PN}{pseudo-noise}
\acro{PPM}{pulse position modulation}
\acro{PRake}{Partial Rake}
\acro{PSD}{power spectral density}
\acro{PSEP}{pairwise synchronization error probability}
\acro{PSK}{phase shift keying}
\acro{PD}{power density}
\acro{8-PSK}[$8$-PSK]{$8$-phase shift keying}
\acro{PPP}{Poisson point process}
\acro{PCP}{Poisson cluster process}
 
\acro{FSK}{Frequency Shift Keying}

\acro{QAM}{Quadrature Amplitude Modulation}
\acro{QPSK}{Quadrature Phase Shift Keying}
\acro{OQPSK/PM}{OQPSK with phase modulator }

\acro{RD}[RD]{raw data}
\acro{RDL}{"random data limit"}
\acro{ric}[RIC]{restricted isometry constant}
\acro{rict}[RICt]{restricted isometry constant threshold}
\acro{rip}[RIP]{restricted isometry property}
\acro{ROC}{receiver operating characteristic}
\acro{rq}[RQ]{Raleigh quotient}
\acro{RS}[RS]{Reed-Solomon}
\acro{RSC}[RSSC]{RS based source coding}
\acro{r.v.}{random variable}                               
\acro{R.V.}{random vector}
\acro{RMS}{root mean square}
\acro{RFR}{radiofrequency radiation}
\acro{RIS}{Reconfigurable Intelligent Surface}
\acro{RNA}{RiboNucleic Acid}
\acro{RRM}{Radio Resource Management}
\acro{RUE}{reference user equipments}
\acro{RAT}{radio access technology}
\acro{RB}{resource block}

\acro{SA}[SA-Music]{subspace-augmented MUSIC with OSMP}
\acro{SC}{small cell}
\acro{SCBSES}[SCBSES]{Source Compression Based Syndrome Encoding Scheme}
\acro{SCM}{sample covariance matrix}
\acro{SEP}{symbol error probability}
\acro{SG}[SG]{sparse-land Gaussian model}
\acro{SIMO}{single-input multiple-output}
\acro{SINR}{signal-to-interference plus noise ratio}
\acro{SIR}{signal-to-interference ratio}
\acro{SISO}{Single-Input Single-Output}
\acro{SMV}{single measurement vector}
\acro{SNR}[\textrm{SNR}]{signal-to-noise ratio} 
\acro{sp}[SP]{subspace pursuit}
\acro{SS}{spread spectrum}
\acro{SW}{sync word}
\acro{SAR}{specific absorption rate}
\acro{SSB}{synchronization signal block}
\acro{SR}{shrink and realign}

\acro{tUAV}{tethered Unmanned Aerial Vehicle}
\acro{TBS}{terrestrial base station}

\acro{uUAV}{untethered Unmanned Aerial Vehicle}
\acro{PDF}{probability density functions}

\acro{PL}{path-loss}

\acro{TH}{time-hopping}
\acro{ToA}{time-of-arrival}
\acro{TR}{transmitted-reference}
\acro{TW}{Tracy-Widom}
\acro{TWDT}{TW Distribution Tail}
\acro{TCM}{trellis coded modulation}
\acro{TDD}{Time-Division Duplexing}
\acro{TDMA}{Time Division Multiple Access}
\acro{Tx}{average transmit}

\acro{UAV}{Unmanned Aerial Vehicle}
\acro{uric}[URIC]{upper restricted isometry constant}
\acro{urict}[URICt]{upper restricted isometry constant threshold}
\acro{UWB}{ultrawide band}
\acro{UWBcap}[UWB]{Ultrawide band}   
\acro{URLLC}{Ultra Reliable Low Latency Communications}
         
\acro{wuric}[UWRIC]{upper weak restricted isometry constant}
\acro{wurict}[UWRICt]{UWRIC threshold}                
\acro{UE}{User Equipment}
\acro{UL}{UpLink}

\acro{WiM}[WiM]{weigh-in-motion}
\acro{WLAN}{wireless local area network}
\acro{wm}[WM]{Wishart matrix}                               
\acroplural{wm}[WM]{Wishart matrices}
\acro{WMAN}{wireless metropolitan area network}
\acro{WPAN}{wireless personal area network}
\acro{wric}[WRIC]{weak restricted isometry constant}
\acro{wrict}[WRICt]{weak restricted isometry constant thresholds}
\acro{wrip}[WRIP]{weak restricted isometry property}
\acro{WSN}{wireless sensor network}                        
\acro{WSS}{Wide-Sense Stationary}
\acro{WHO}{World Health Organization}
\acro{Wi-Fi}{Wireless Fidelity}

\acro{sss}[SpaSoSEnc]{sparse source syndrome encoding}

\acro{VLC}{Visible Light Communication}
\acro{VPN}{Virtual Private Network} 
\acro{RF}{Radio Frequency}
\acro{FSO}{Free Space Optics}
\acro{IoST}{Internet of Space Things}

\acro{GSM}{Global System for Mobile Communications}
\acro{2G}{Second-generation cellular network}
\acro{3G}{Third-generation cellular network}
\acro{4G}{Fourth-generation cellular network}
\acro{5G}{Fifth-generation cellular network}	
\acro{gNB}{next-generation Node-B Base Station}
\acro{NR}{New Radio}
\acro{UMTS}{Universal Mobile Telecommunications Service}
\acro{LTE}{Long Term Evolution}

\acro{QoS}{Quality of Service}
\end{acronym}
	
\newcommand{\SAR} {\mathrm{SAR}}
\newcommand{\WBSAR} {\mathrm{SAR}_{\mathsf{WB}}}
\newcommand{\gSAR} {\mathrm{SAR}_{10\si{\gram}}}
\newcommand{\Sab} {S_{\mathsf{ab}}}
\newcommand{\Eavg} {E_{\mathsf{avg}}}
\newcommand{\ft}{f_{\textsf{th}}}
\newcommand{\alphatf}{\alpha_{24}}

\title{
End-to-End Uplink Performance Analysis of Satellite-Based IoT Networks: A Stochastic Geometry Approach
}


\author{
Jiusi Zhou, Ruibo Wang, Basem Shihada, (Senior Member, IEEE), and Mohamed-Slim Alouini, {\em Fellow, IEEE}
\thanks{The authors are with King Abdullah University of Science and Technology (KAUST), CEMSE division, Thuwal 23955-6900, Saudi Arabia. Corresponding author: Ruibo Wang. (e-mail: jiusi.zhou@kaust.edu.sa; ruibo.wang@kaust.edu.sa; basem.shihada@kaust.edu.sa; slim.alouini@kaust.edu.sa).  
}
\vspace{-6mm}
}
\maketitle
\thispagestyle{empty}

\begin{abstract}
With the deployment of satellite constellations, Internet-of-Things (IoT) devices in remote areas have gained access to low-cost network connectivity. In this paper, we investigate the performance of IoT devices connecting in up-link through low Earth orbit (LEO) satellites to geosynchronous equatorial orbit (GEO) links. We model the dynamic LEO satellite constellation using the stochastic geometry method and provide an analysis of end-to-end availability with low-complexity and coverage performance estimates for the mentioned link. Based on the analytical expressions derived in this research, we make a sound investigation on the impact of constellation configuration, transmission power, and the relative positions of IoT devices and GEO satellites on end-to-end performance.
\end{abstract}

\maketitle

\section{Introduction}
{\color{black}
Recently, non-terrestrial networks have garnered significant attention due to their seamless global coverage \cite{10520169}, enabling even remote Internet-of-Things (IoT) devices to connect to the network \cite{lin2021supporting,lin2021secrecy}. Remote areas usually lack sufficient terrestrial communication infrastructure, such as fiber optic networks or cellular towers. Therefore, IoT devices in many cases can only access the network via satellite or other aerial links \cite{fraire2022space}. However, the power sources for IoT devices are typically limited, relying on batteries, solar energy, or wireless charging technology \cite{liu2019toward,kishk2016downlink}. This implies that the signal transmission power of IoT devices is ultimately weaker \cite{10185203}. Facing substantial path loss in long-distance communications \cite{kodheli2021nb}, for instance, in direct transmission with geosynchronous equatorial orbit (GEO) satellites, IoT devices need to rely on alternative means to maintain stable communication. The research has demonstrated the feasibility of direct communication between IoT devices and low Earth orbit (LEO) satellites \cite{qu2017leo}. Therefore, establishing an IoT device to GEO satellite link with LEO satellites as relays is a reasonable approach \cite{huang2018online}. }

\begin{figure}[htbp]
\centering
\includegraphics[width=0.7\linewidth]{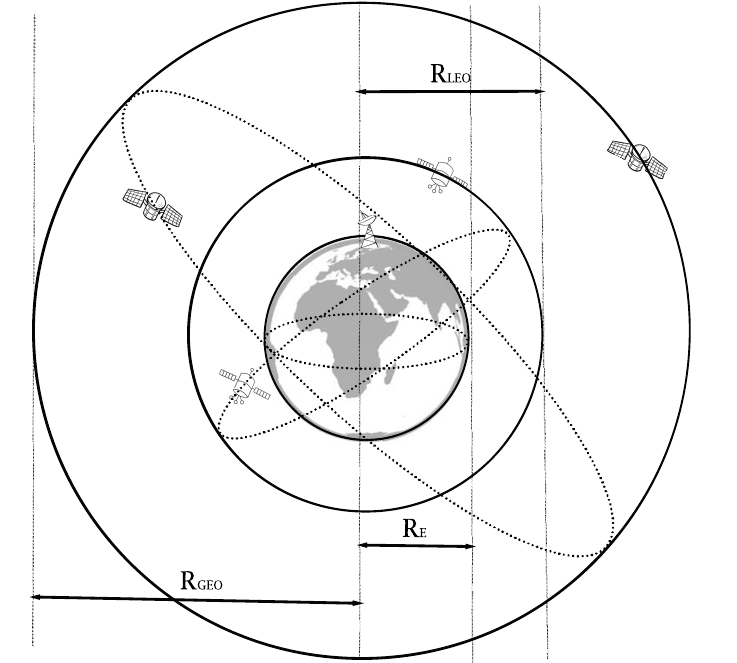}
\caption{Diagram of satellite-based IoT networks.}
\label{sys1}
\end{figure}

\par
For the uplink IoT device-LEO satellite-GEO satellite communication link, analyzing its end-to-end performance is a critical and meaningful research topic \cite{dwivedi2020performance}. Performance analysis allows us to know the probability of connection for a given transmission power from IoT devices. These insights aid in better energy management and allocation for IoT devices. However, due to the non-geostationary nature of LEO satellites, accurately modeling and analyzing the aforementioned end-to-end communication link becomes challenging \cite{wang2023reliability}.

So far, the vast majority of literature achieves the dynamic variation of LEO satellite positions through numerical simulations \cite{lim2020performance, chen2023remote}. However, modeling and simulating massive LEO satellite constellations are computationally expensive. When system parameters, such as the transmission power of IoT devices, are adjusted, simulations have to be re-executed. Therefore, we aim to represent metrics such as coverage probability as functions of system parameters like transmission power through desired analytical methods, reducing computational complexity. In the literature providing analytical results for air-to-ground communication systems, very few studies have considered the randomness of the positions of GEO and LEO satellites in the satellite-relayed communication system \cite{lin2023leo}. Although the authors in \cite{lin2023leo, ma2022secure, zhang2022outage} considered the randomness in the distribution of satellites or other aerial devices, they did not model the satellite constellation exactly. As a result, they only accounted for the dynamic variation of satellite positions within a given orbit, lacking a comprehensive view of the entire satellite system's performance.

\subsection{Related Works}
Based on the above discussion, we introduce stochastic geometry (SG) as a mathematical tool suitable for dynamic network performance analysis \cite{lou2023coverage}. {\color{black}Under the SG framework, LEO satellites are modeled as a stochastic point process and the spherical binomial point process (BPP) is one of the most common models among them \cite{talgat2020stochastic,lou2023haps}. The core idea of the SG analytical framework is to achieve strong tractability by assuming that satellites follow a specific distribution. Since the BPP model does not account for the modeling of orbits, it differs from actual satellite constellation models. Fortunately, both in terms of network topology and performance estimation outcomes, the differences between BPP and deterministic constellations are neglected \cite{wang2022evaluating, ok-1}. Therefore, although the spatial distribution of the satellite constellation is regular but not completely random, we can still model satellite constellations as a BPP, accepting a small cost and reasonable modeling differences in exchange for the strong tractability it provides.}

\begin{figure}[htbp]
\centering
\includegraphics[width=0.7\linewidth]{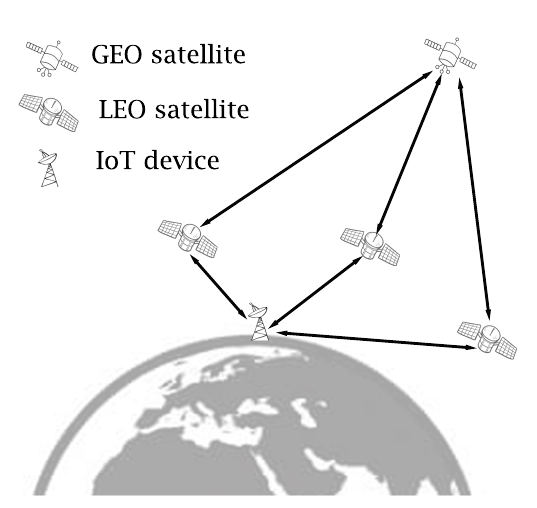}
\caption{Diagram of satellite-based IoT networks.}
\label{sys2}
\end{figure}

\par
{\color{black}There are several SG-based studies analyzing the performance of satellite networks under the stochastic geometry framework. The authors in \cite{chan2022stochastic} considered the uplink performance analysis of IoT-over-satellite link. The above satellite-based IoT network coverage analysis framework is further expanded in \cite{10463093,hong2024narrowband}. Specifically, Authors in \cite{10463093} adopted a more practical modeling method, and authors in \cite{hong2024narrowband} proposed an adaptive coverage enhancement strategy.} However, research on satellite relay systems remains limited. This is due to the complex distance distribution in spherical multi-hop networks, posing significant challenges in providing low-complexity analytical results. {\color{black}To our best knowledge, within the SG framework, only our previous research \cite{wang2022conditional} has analyzed the distance distribution of satellite-relay systems. In this study, we considered a dual-hop communication scenario where signals are transmitted from a ground transmitter to a ground receiver via an LEO satellite as the relay. We respectively derived the analytical expressions for the distance distribution of each hop. However, our study did not involve performance analysis nor encompass modeling of GEO satellites or channel modeling, thus differing significantly from the scope of this article.}

\subsection{Contribution}
As the first study to analyze the performance of IoT-GEO satellite link's performance under the SG framework, the contributions of this article are as follows.

\begin{itemize}
\item {\color{black} We derive the distance distributions of the IoT-LEO satellite link, as well as the LEO satellite-GEO satellite link (LGL).} Due to the non-independence of the distance distributions between the two links, the distance distribution derivation in this paper differs significantly from existing studies based on spherical SG.
\item For the first time, we incorporate the inter-satellite link channel model into the modeling of SG-based satellite coverage issues. Taking pointing errors into account as small-scale fading, we analyze the coverage performance of the LGL.
\item Unlike most SG-based studies on satellite networks focusing on point-to-point performance, we analyze the end-to-end performance of the dual-hop link. Specifically, we first derive the probability of having at least one LEO satellite being within the common reliable communication range of the IoT device and the GEO satellite, called the availability probability. We derive the analytical expression for coverage probability, which is the probability of end-to-end successful communication given that the availability is ensured. 
\item Through numerical simulations, we validate the accuracy performance of the aforementioned analytical expressions and present the impact of constellation configuration, transmission power, and the relative positioning of IoT devices and GEO satellites, on these metrics. 
\end{itemize}

\section{System Model}
In this section, we establish the spatial model of the IoT device, GEO satellite, and LEO satellite constellation. Then, channel models of the IoT device-LEO satellite link (ILL) and LGL are provided.

\subsection{Spatial Model}
We consider a system including an IoT device, a GEO Satellite, and $N_{\mathrm{LEO}}$ LEO satellites. LEO satellites are distributed on a sphere around Earth with a radius $R_{\mathrm{LEO}}=R_{\oplus}+h_{\mathrm{LEO}}$ form a spherical homogeneous BPP, where $R_{\oplus}$ is the radius of the Earth and $h_{\mathrm{LEO}}$ denotes the altitude of LEO satellites. According to Slivnyak's theorem \cite{feller1991introduction}, the rotation of the coordinate system does not affect the distribution of homogeneous point processes. Without loss of generality, we take the Earth's center as the origin and establish a spherical coordinate system. The coordinates of the IoT device and the GEO satellites are denoted as $x_{\mathrm{IoT}} (R_{\oplus},0,0)$ and $x_{\mathrm{GEO}} (R_{\mathrm{GEO}},\Theta,0)$, where $R_{\mathrm{GEO}} = R_{\oplus}+h_{\mathrm{GEO}}$ is the radius of the sphere that the GEO satellite is located. Here $\Theta$ represents the central angle between the IoT device and the GEO satellite, and the definition of the central angle is given as follows.

\begin{definition}[Central Angle]
    The central angle between two communication devices, A and B, refers to the angle created by the lines extending from the Earth's center to device A and device B respectively \cite{wang2022ultra}.
\end{definition}

\subsection{Channel Model}
In this article, we consider the uplink channel models to follow the free space propagation model experienced with large-scale fading and small-scale fading, and the received power is given as \cite{na2021performance}
\begin{equation}
    \rho_Q^r = \rho_Q^t G_Q \left( \frac{\lambda_Q}{4\pi l_Q} \right)^2 \zeta_Q W_Q, \ l_Q \leq l_Q^{\max},
\end{equation}
where $\rho_Q^t$, $G_Q$, $\lambda_Q$, $\zeta_Q$, and $W_Q$ denote the transmission power, antenna gain, wavelength, additional attenuation during propagation, and the power of small-scale fading, respectively; $Q \in \{{\mathrm{IL},{\mathrm{LG}}}\} $. Parameters with subscripts ${\mathrm{IL}}$ and  ${\mathrm{LG}}$ represent labels for the two links, ILLs and LGLs respectively. We consider $\zeta_{\mathrm{IL}}$ is mainly caused by rain attenuation \cite{talgat2020stochastic} and the attenuation of satellite-GEO satellite link is negligible ($\zeta_{\mathrm{LG}} = 0$~dB).

\par
Considering the energy constraints of the IoT device, we assume that the IoT device will select the LEO satellite closest to it as the relay. $l_{\mathrm{IL}}$ and $l_{\mathrm{LG}}$ denote the distance between the IoT device and the selected relay LEO satellite, and the distance between this LEO satellite and the GEO satellite. Furthermore, $l_{\mathrm{IL}}^{\max}$ and $l_{\mathrm{LG}}^{\max}$ are the maximum distances that can maintain stable communication for ILLs and LGLs. To ensure these links are not blocked by the Earth, $l_{\mathrm{IL}}^{\max} \leq \sqrt{R_{\mathrm{LEO}}^2 - R_{\oplus}^2}$ and $l_{\mathrm{LG}}^{\max} \leq 2\sqrt{R_{\mathrm{GEO}}^2 - R_{\mathrm{LEO}}^2}$ need to be satisfied.

\par
The small-scale fading of ILL is assumed to follow the shadowed-Rician (SR) fading. Considering the shadowing and multi-path effects, SR fading stands out as one of the most precise models describing small-scale fading in space-terrestrial communication links. The cumulative distribution function (CDF) of the SR fading power $W_{\mathrm{IL}}$ is given as follows \cite{abdi2003new}: 
\begin{equation}\label{SR}
\begin{split}
    F_{W_{\mathrm{IL}}} (w) & = \left( \frac{2 b_0 m}{2 b_0 m + \Omega} \right)^{m} \sum_{z=0}^{\infty} \frac{(m)_z}{z! \Gamma(z+1)} \\
    & \times \left( \frac{\Omega}{2 b_0 m + \Omega} \right)^z \Gamma_l\left(z+1, \frac{w}{2b_0} \right),
\end{split}
\end{equation}
where $(m)_z$ is the Pochhammer symbol, $m$, $b_0$ and $\Omega$ are parameters of the SR fading.  $\Gamma( \cdot )$ and $\Gamma_l(\cdot, \cdot )$ denote the gamma function and lower incomplete gamma function, respectively.

\par
Furthermore, the shadowing and multi-path effects have minimal impact in the space \cite{gopal2014modulation}. In this case, the pointing error becomes the primary factor in small-scale fading due to the rapid movement of LEO satellites. Therefore, we consider $W_{\mathrm{LG}}$ follows the pointing error model  \cite{ata2022performance}. Given that the deviation angle of the beam as $\theta_d$, the conditional probability density function (PDF) of $W_{\mathrm{LG}}$ can be written as 
\begin{equation}
\label{pointing_error}
    f_{W_{\mathrm{LG}} \, | \, \theta_d}\left ( w \right ) = \frac{\eta_s^2 w^{\eta_s^2-1 } \cos\left ( \theta_d \right )}{A_0^{\eta_s^2}},  \ 0 \leq w \leq A_0,
\end{equation}
where $\eta_s$ and $A_0$ are parameters of the pointing error, and the deviation angle $\theta_d$ is subject to Rayleigh distribution with variance  $\varsigma^2$\cite{ata2022performance}, 
\begin{equation}
    f_{\theta_d}\left ( \theta_d \right )=\frac{\theta_d}{\varsigma^2}\exp\left ( -\frac{\theta_d^2}{2\varsigma^2} \right ), \ \theta_d\geq 0.
\end{equation}

\section{Performance Analysis}
In this subsection, we give the definitions and analytical expressions of two end-to-end performance metrics, namely satellite availability probability and coverage probability in order.

\subsection{Satellite Availability}
In the system model, both ILL and LGL have been set with upper bounds for reliable communication distances. {\color{black}Therefore, the presence of a relay LEO satellite within the common reliable communication range of IoT devices and GEO satellites is a prerequisite for communication. If the required above conditions are achieved, the LEO satellite is considered available.}

\begin{definition}[Satellite Availability Probability]
The satellite availability probability is the probability that the selected relay LEO satellite's distance to the IoT device is less than $l_{\mathrm{IL}}^{\max}$ and its distance to the GEO satellite is less than $l_{\mathrm{LG}}^{\max}$.
\end{definition}

{\color{black}
Because the closest LEO satellite is selected as a relay to the IoT device and the availability of LGL depends on the relay satellite's position, we present the following lemma for the PDF of the central angle in our IoT-LEO model regarding the distance between the IoT and the relay satellite.} As we have employed a spherical coordinate system for modeling, expressing distance in terms of central angles proves to be more concise.

\begin{lemma}\label{lemma1}
The PDF of the central angle between the IoT device and the LEO satellite is
\begin{equation}\label{contact}
    f_{\theta_{\mathrm{IL}}}(\theta) = \frac{N_{\mathrm{LEO}} \sin\theta}{2} \left( 
    \frac{1+\cos\theta}{2} \right)^{N_{\mathrm{LEO}}-1},
\end{equation}
where $N_{\mathrm{LEO}}$ represents the number of LEO satellites, $\theta \leq \theta_{\mathrm{IL}}^{\max}$, and $\theta_{\mathrm{IL}}^{\max}$ can be expressed as,
\begin{equation}\label{thetamax}
    \theta_{\mathrm{IL}}^{\max} = \arccos \left( \frac{R_{\mathrm{LEO}}^2 + R_{\oplus}^2 - \left(l_{\mathrm{LEO}}^{\max}\right)^2 }{2 R_{\mathrm{LEO}} R_{\oplus} } \right).
\end{equation}
\begin{IEEEproof}
    See Appendix~\ref{app:lemma1}
\end{IEEEproof}
\end{lemma}

Based on the above central angle distribution, also known as the contact angle distribution \cite{wang2022stochastic}, the satellite availability probability can be determined by the following theorem.

\begin{theorem}\label{theorem1}
The LEO satellite availability probability can be given by the following expression
\begin{equation}
P^A = \int_0^{ \theta_{\mathrm{IL}}^{\max} } f_{\theta_{\mathrm{IL}}}(\theta)  P_{\mathrm{LG}}^A (\theta) {\mathrm{d}}\theta,
\end{equation}
where $f_{\theta_{\mathrm{IL}}}(\theta)$ and $\theta_{\mathrm{IL}}^{\max}$ are defined in (\ref{contact}) and (\ref{thetamax}), respectively. $P_{\mathrm{LG}}^A (\theta)$ is called the conditional availability probability of LGL, given that the contact angle is $\theta$, and its expression is given as
\begin{align}\label{PALG}
P_{\mathrm{LG}}^A (\theta) = \left\{
\begin{array}{lll}
0, \ \ \ \ \sqrt{ R_{\mathrm{LEO}}^2 + R_{\mathrm{GEO}}^2 - 2 R_{\mathrm{LEO}} R_{\mathrm{GEO}} (\sin\theta \sin\Theta + \cos\theta \cos\Theta )} > l_{\mathrm{LG}}^{\max}, \\
1, \ \ \ \ \sqrt{ R_{\mathrm{LEO}}^2 + R_{\mathrm{GEO}}^2 - 2 R_{\mathrm{LEO}} R_{\mathrm{GEO}} ( \cos\theta \cos\Theta - \sin\theta \sin\Theta)} < l_{\mathrm{LG}}^{\max}, \\
\frac{1}{\pi} \arccos\left( \frac{R_{\mathrm{LEO}}^2 + R_{\mathrm{GEO}}^2 - \left( l_{\mathrm{LG}}^{\max} \right)^2 }{2 R_{\mathrm{LEO}} R_{\mathrm{GEO}} \sin\theta \sin\Theta} - \frac{\cos\theta \cos\Theta}{\sin\theta \sin\Theta} \right), \ \ \ \ \ \ \ \ \ \ \ \ \ \ \ \ \ \ {\mathrm{otherwise}}.
\end{array}
\right.
\end{align}
\begin{IEEEproof}
    See Appendix~\ref{app:theorem1}.
\end{IEEEproof}
\end{theorem}

As the theorem demands specific requirements for the distances of both links, thus it can be considered an end-to-end performance metric. {\color{black}Considering that the availability probabilities of LGL and dual-hop have been analyzed, we can derive the following corollary about the availability probability of ILL based on the above theorem.}

\begin{corollary}
The availability probability of ILL, which is the probability that the distance between the LEO satellite and the IoT device is less than $\theta_{\mathrm{IL}}^{\max}$, is given by
\begin{equation}
\begin{split}
    P_{\mathrm{IL}}^A = 1 - \left( \frac{ 1 + \cos\theta_{\mathrm{IL}}^{\max} }{2} \right)^{N_{\mathrm{LEO}}},
\end{split}
\end{equation}
where $\theta_{\mathrm{IL}}^{\max}$ is defined in (\ref{thetamax}).
\end{corollary}

\subsection{Coverage Probability}
In this subsection, we first provide the definition and mathematical expression of coverage probability, followed by the deductions for analytical expression. 

\begin{definition}[Coverage Probability]
The coverage probability is the probability that the signal-to-noise ratio (SNR) of ILL is greater than the coverage threshold $\gamma_{\mathrm{IL}}$ and the SNR of LGL greater than the coverage threshold $\gamma_{\mathrm{LG}}$.
\end{definition}

\par
Mathematically, the coverage probability can be given by
\begin{equation}
    P^C = \mathbbm{P} \left[ \frac{\rho_{\mathrm{IL}}^r}{\sigma_{\mathrm{LEO}}^2} > \gamma_{\mathrm{IL}}, \frac{\rho_{\mathrm{LG}}^r}{\sigma_{\mathrm{GEO}}^2} > \gamma_{\mathrm{LG}} \right],
\end{equation}
where $\sigma_{\mathrm{LEO}}^2$ and $\sigma_{\mathrm{GEO}}^2$ denote the noise power at the LEO satellite and the GEO satellite. {\color{black} Coverage probability is one of the most widely used performance metrics for evaluating network performance. It represents the ability of a communication receiver to successfully demodulate the signal.} Before providing the analytical expression for coverage probability, we need to derive the unconditional CDF of pointing errors as a lemma.

\begin{lemma}\label{lemma2}
The unconditional CDF of pointing errors is approximately given by
\begin{align}\label{point}
F_{W_{\mathrm{LG}}} (w) = \left\{
\begin{array}{lll}
0, & w < 0, \\
\frac{w^{\eta_s^2}}{A_0^{\eta_s^2}} \left( 1 - \varsigma^2 \right) & 0 \leq w \leq A_0, \\
1, & w > A_0. \\
\end{array}
\right.
\end{align}
\begin{IEEEproof}
    See Appendix~\ref{app:lemma2}.
\end{IEEEproof}
\end{lemma}

Based on the above lemma, the end-to-end coverage probability can be deduced by the following theorem.

\begin{theorem}\label{theorem2}
The coverage probability is
given in as
\begin{equation}\label{cover}
\begin{split}
    & P^C = \int_0^{ \theta_{\mathrm{IL}}^{\max}} \int_0^{ P_{\mathrm{LG}}^A(\theta) } \left( 1 - F_{W_{\mathrm{IL}}} \left( \frac{\gamma_{\mathrm{IL}} \sigma_{\mathrm{LEO}}^2 }{\rho_{\mathrm{IL}}^t G_{\mathrm{IL}} \zeta_{\mathrm{IL}} } \left( \frac{4 \pi}{\lambda_{\mathrm{IL}}} \right)^2 \left( R_{\mathrm{LEO}}^2 + R_{\oplus}^2 - 2R_{\mathrm{LEO}} R_{\oplus} \cos\theta \right) \right) \right) \\
    & \times \left( 1 - F_{W_{\mathrm{LG}}} \left( \frac{\gamma_{\mathrm{LG}} \sigma_{\mathrm{GEO}}^2 }{\rho_{\mathrm{LG}}^t G_{\mathrm{LG}} \zeta_{\mathrm{LG}} } \left( \frac{4 \pi}{\lambda_{\mathrm{LG}}} \right)^2 d^2 \left( R_{\mathrm{GEO}},\Theta,0; R_{\mathrm{LEO}},\theta,\varphi \right) \right) \right) \frac{\sin\theta f_{\theta_{\mathrm{IL}}}(\theta) }{4 \pi} {\mathrm{d}}\varphi {\mathrm{d}}\theta.
\end{split}
\end{equation}
where $\theta_{\mathrm{IL}}^{\max}$, $P_{\mathrm{LG}}^A$, $F_{W_{\mathrm{IL}}}$ and $F_{W_{\mathrm{LG}}}$ are defined in (\ref{thetamax}), (\ref{PALG}), (\ref{SR}) and (\ref{point}), respectively. $d \left( R_1, \theta_1, \varphi_1 ; R_2, \theta_2, \varphi_2 \right )$ calculates the Euclidean distance between position $(R_1, \theta_1, \varphi_1)$ and position $(R_2, \theta_2, \varphi_2)$ in the spherical coordinate system,
\begin{sequation}\label{Euclidean}
\begin{split}
    & d \left( R_1, \theta_1, \varphi_1 ; R_2, \theta_2, \varphi_2 \right ) \\
    & = \sqrt{R_1^2 + R_2^2 - 2  R_1 R_2 (\sin\theta_1 \sin\theta_2 \cos(\varphi_1-\varphi_2) + \cos\theta_1 \cos\theta_2 )}.
\end{split}
\end{sequation}
\begin{IEEEproof}
See Appendix~\ref{app:theorem2}.
\end{IEEEproof}
\end{theorem}

\par
It can be shown that coverage probability is determined by the SNR of two links, thus it is also an end-to-end performance metric.

\begin{table*}[]
\centering
\caption{Simulation Parameters \cite{talgat2020stochastic,ata2022performance}.}
\label{table1}
\resizebox{\linewidth}{!}{ 
\renewcommand{\arraystretch}{1.1}
\begin{tabular}{|c|c|c|}
\hline
Notation     & Meaning                         & Default Value      \\ \hline \hline
$N_{\mathrm{LEO}}$ & Number of LEO satellites     & 1000  \\ \hline
$R_{\oplus},R_{\mathrm{LEO}},R_{\mathrm{GEO}}$ & Radius of the Earth, LEO satellites, and GEO satellites  & $6371,7371,35860$~km  \\ \hline
$\sigma_{\mathrm{LEO}}^2, \sigma_{\mathrm{GEO}}^2$ & Noise power  & $5 \times 10^{-10}$~mW \\ \hline
$\Omega,b_0,m$  &  Parameters of the SR fading  &  $1.29,0.158,19.4$ \\ \hline
$\eta_s$, $A_0$ & Parameters of the pointing error                  & $1.00526,3.2120$ 
\\ \hline
$\varsigma$  & Variance of Rayleigh distribution & $15$~mrad \\ \hline
$\zeta_{\mathrm{IL}}$, $\zeta_{\mathrm{LG}}$ &  Additional attenuation during
propagation & $-2$~dB, $0$~dB \\ \hline
$G_{\mathrm{ST}}, G_{\mathrm{SS}}$ & Antenna gain   & $41.7$~dBi           \\ \hline
$\lambda_{\mathrm{IL}}, \lambda_{\mathrm{LG}}$  & Wavelength  & $1550$~nm   \\ \hline
$\Theta$   & Central angle between the IoT device and the GEO satellite   & $\pi/4$  \\ \hline
$l_{\mathrm{IL}}^{\max},l_{\mathrm{LG}}^{\max}$   & Maximum distance of communication   & $3000$~km, $35000$~km
\\ \hline $\rho_{\mathrm{IL}}^t,\rho_{\mathrm{LG}}^t$ & Transmission power & $15$~dBW, $50$~dBW \\ \hline
\end{tabular}
}
\end{table*}

\section{Numerical Results}
This section demonstrates the numerical results of availability probability and coverage probability. The alignment of the results obtained from Monte Carlo simulations (lines) with the derived analytical results (marks) in the figures confirms the accuracy of the analytical expressions presented in this article. Unless otherwise specified, the parameters in this section will be set to their default values given in Table~\ref{table1}.

\begin{figure}[ht]
\centering
\includegraphics[width=0.7\linewidth]{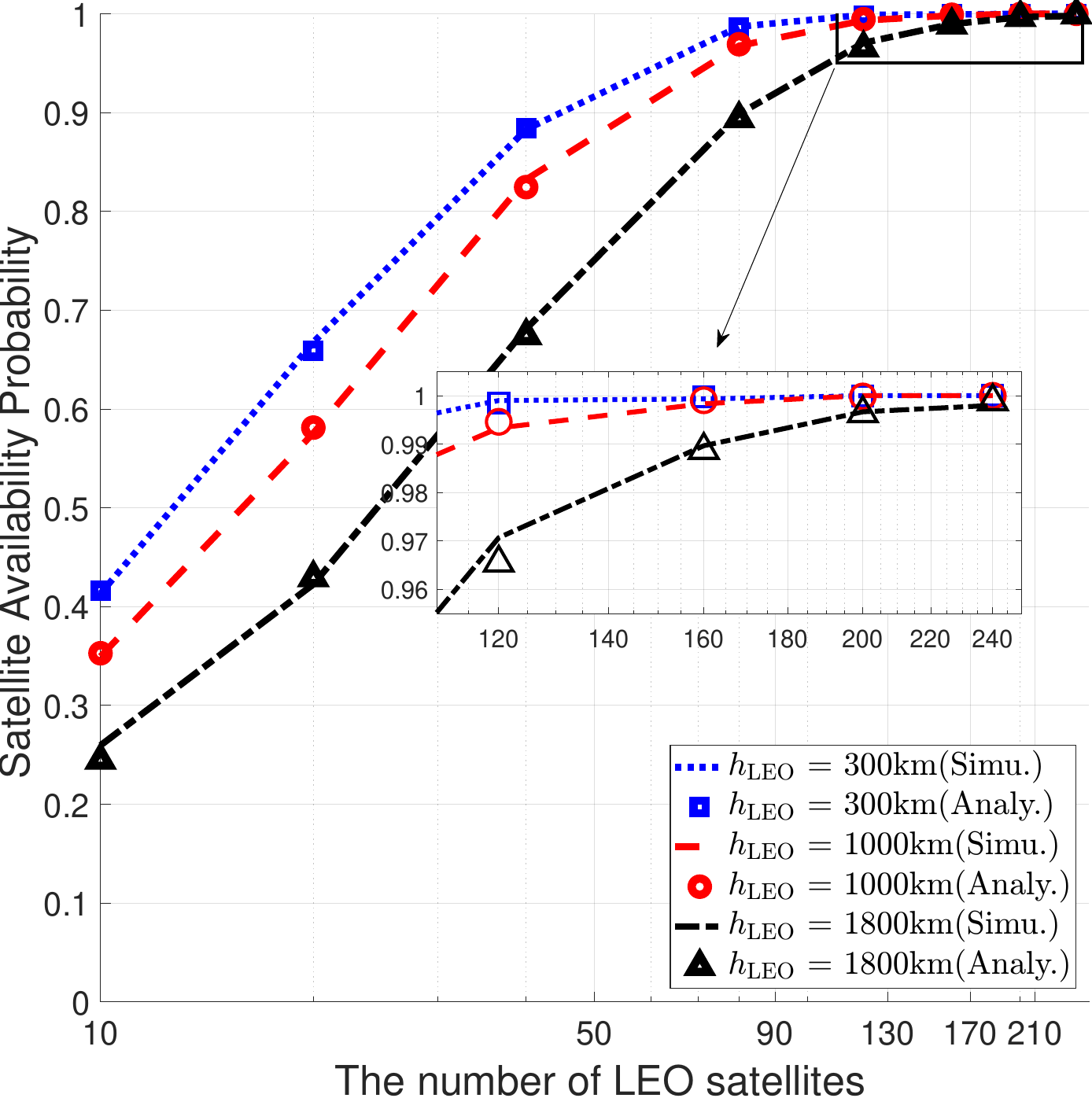}
\caption{Satellite availability probability with different constellation configurations.}
\label{figure2}
\end{figure}

\begin{figure}[htbp]
\centering
\includegraphics[width=0.7\linewidth]{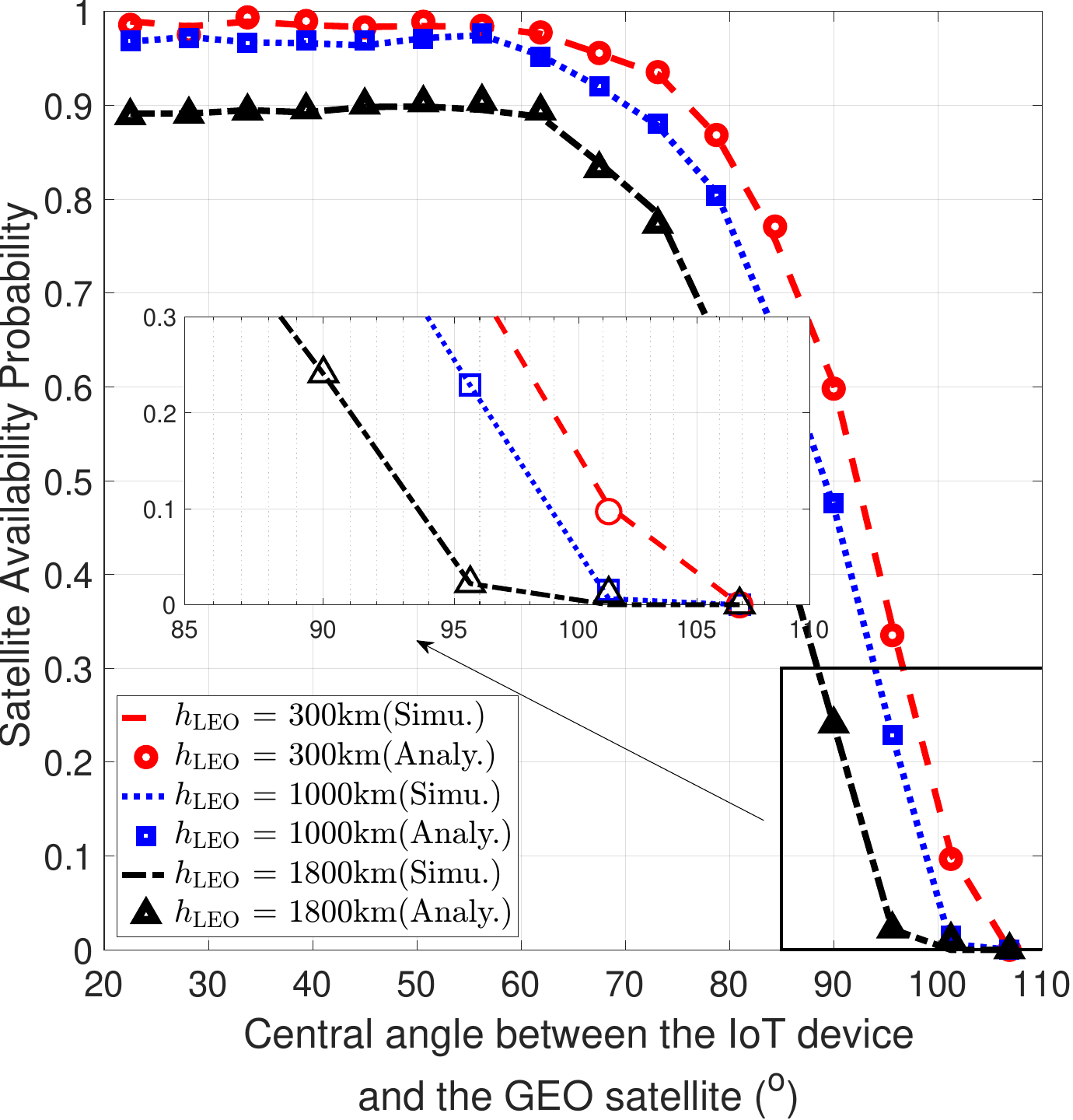}
\caption{Satellite availability probability with different central angles.}
\label{figure3}
\end{figure}

In Fig.~\ref{figure2}, we analyze the impact of the number of LEO satellites and constellation altitude on the availability. {\color{black}For the same number of satellites, constellations at lower altitudes have a larger availability probability.} In the subgraph, we show the number of satellites required to achieve an availability probability of $1$ at different altitudes. In Fig.~\ref{figure3}, the number of satellites is fixed at $100$. When the central angle between IoT devices and GEO satellites is less than $60$ degrees, the influence of the central angle on the availability can be neglected. At this point, availability probability is mainly determined by the ILL. When the central angle is larger than $60$ degrees, the impact of central angles on satellite availability probability is much larger compared to the constellation altitude. The availability probability decreases rapidly as the central angle further increases. 

\begin{figure}[ht]
\centering
\includegraphics[width=0.7\linewidth]{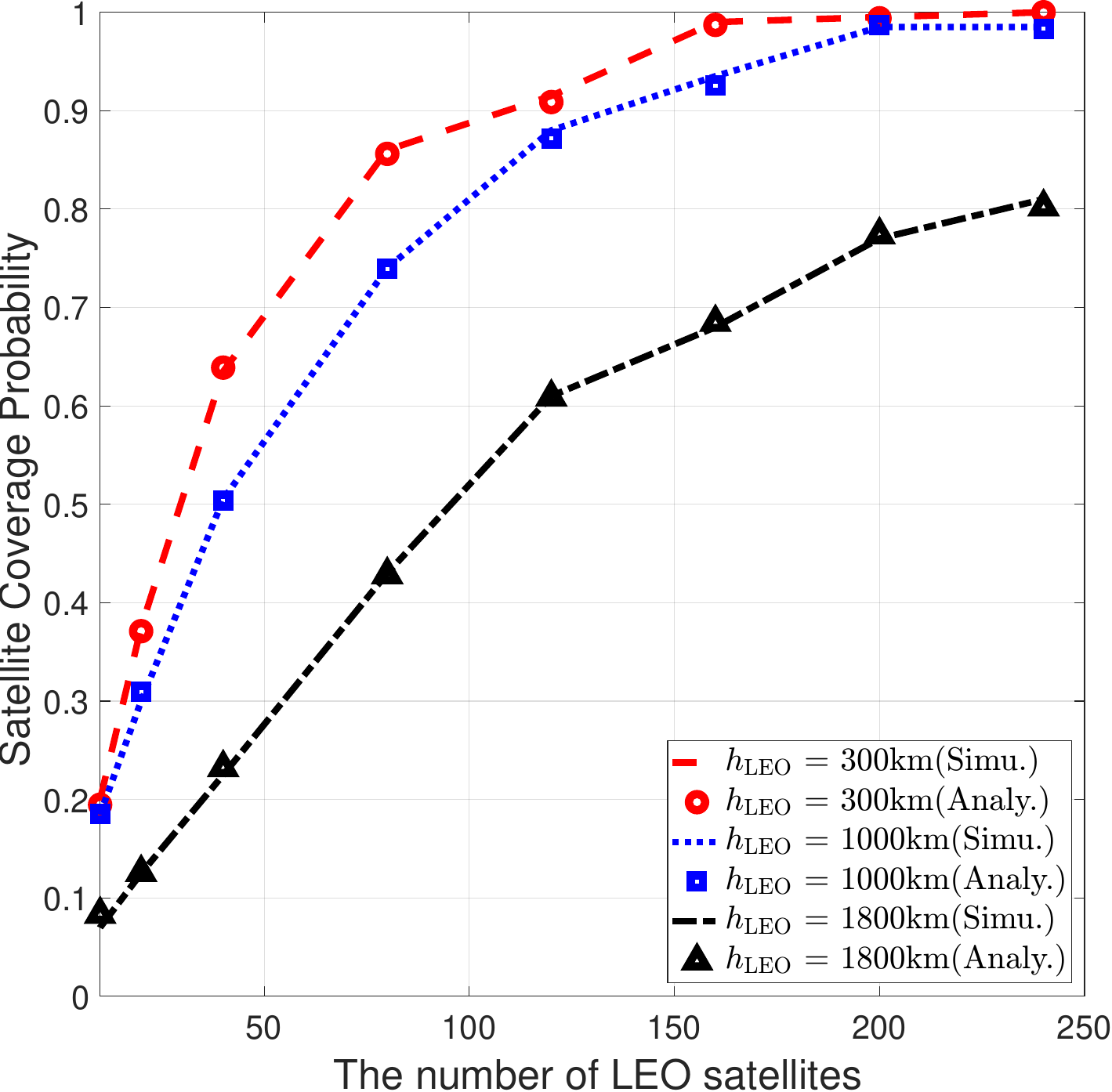}
\caption{Satellite coverage probability with different constellation configurations.}
\label{figure2_2}
\end{figure}

Fig.~\ref{figure2_2} shows the results of satellite coverage probability with different constellation configurations. The lower altitude of LEO satellites and the greater number of satellites result in shorter distances for ILL, thereby improving coverage performance. Under default parameters, compared to LGL, the quality of ILL plays a more decisive role in coverage probability.

\begin{figure}[htbp]
\centering
\includegraphics[width=0.7\linewidth]{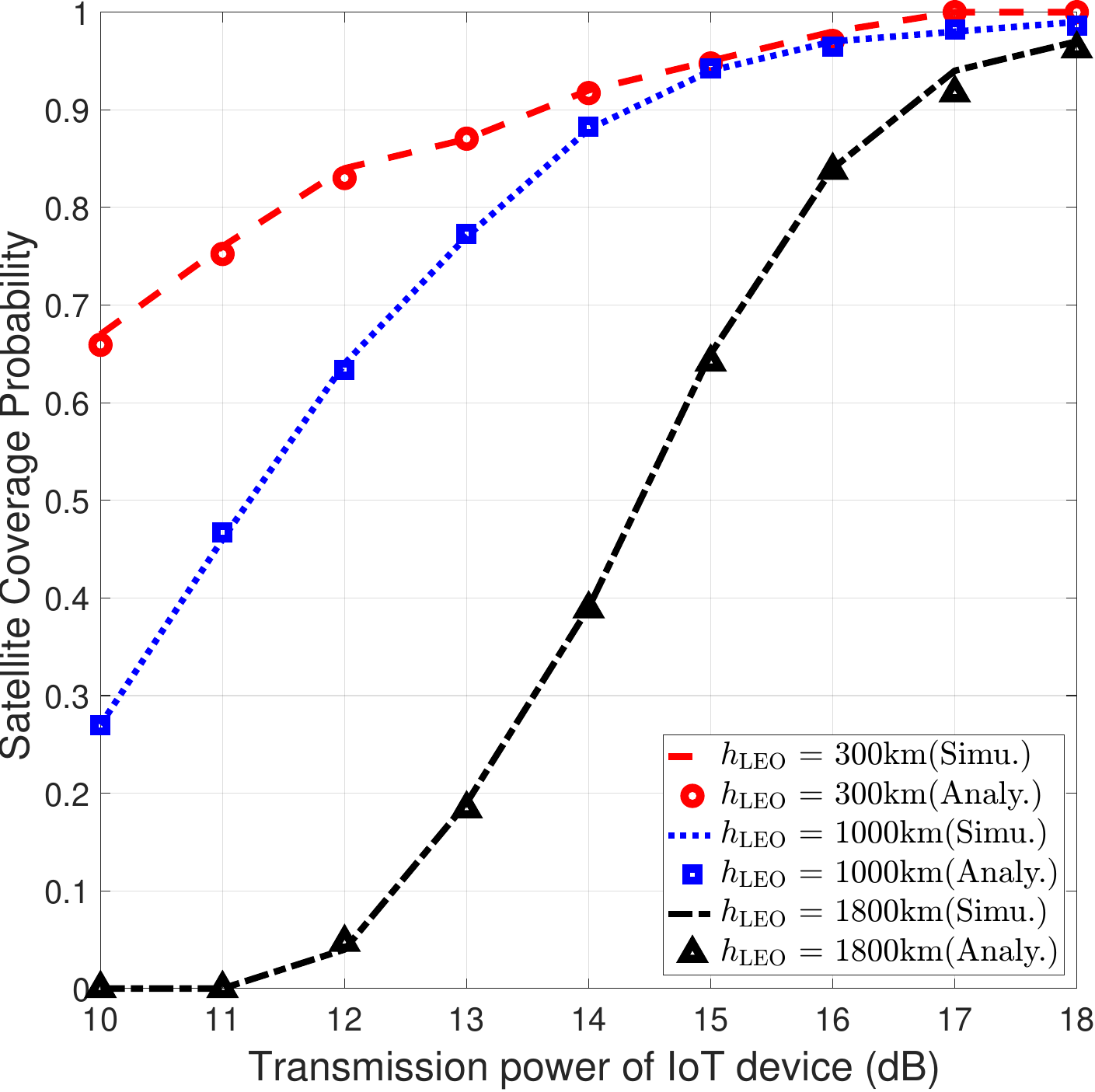}
\caption{Satellite coverage probability with different transmission power of IoT device.}
\label{figure4}
\end{figure}

\begin{figure}[htbp]
\centering
\includegraphics[width=0.7\linewidth]{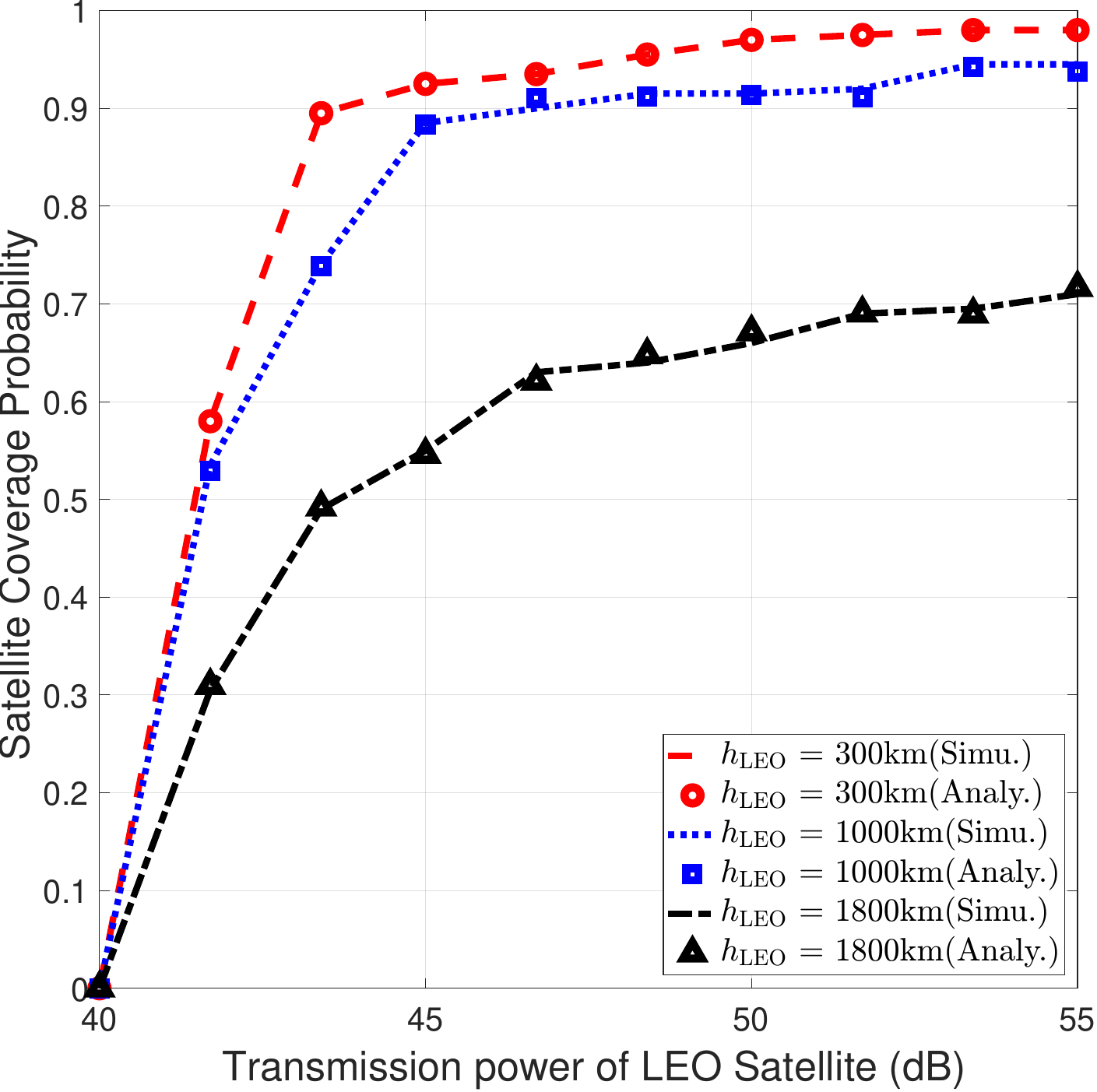}
\caption{Satellite coverage probability with different transmission power of LEO satellite.}
\label{figure5}
\end{figure}

Fig.~\ref{figure4} and Fig.~\ref{figure5} present end-to-end satellite coverage probability with different transmission power of IoT devices and LEO satellites. In general, constellations with lower altitudes have better coverage performance. As shown in Fig.~\ref{figure5}, with the increase in LEO transmission power, the end-to-end coverage probability first increases and then converges. The final converged value depends on the coverage probability of the ILL. 

{\color{black}
\section{Additional Insights}
In this section, we analyze the influence of transmission policy/association strategy on coverage probability, compare different network architectures, and describe the methods of counteracting Doppler shift.

\subsection{Transmission Policy}
In this subsection, we discuss the impact of the following three different transmission policies on coverage probability:
\begin{itemize}
    \item \textbf{Policy 1} (proposed): Select the nearest LEO satellite to the IoT device for relaying.
    \item \textbf{Policy 2}: Select the nearest LEO satellite to the GEO satellite for relaying.
    \item \textbf{Policy 3}: Select the LEO satellite with the shortest distance to the direct link between the IoT device and the GEO satellite for relaying.
\end{itemize}

Fig.~\ref{figure8} compares the coverage probabilities of the three policies when the central angle between the IoT device and the GEO satellite is fixed at $\Theta=\frac{\pi}{12}$. Regardless of the transmission power of the IoT device, our proposed transmission policy (policy 1) consistently outperforms the other two policies. In terms of coverage probability, the performance of Policy 1 and Policy 3 is similar, with both significantly outperforming Policy 2. 

\begin{figure}[htbp]
\centering
\includegraphics[width=0.7\linewidth]{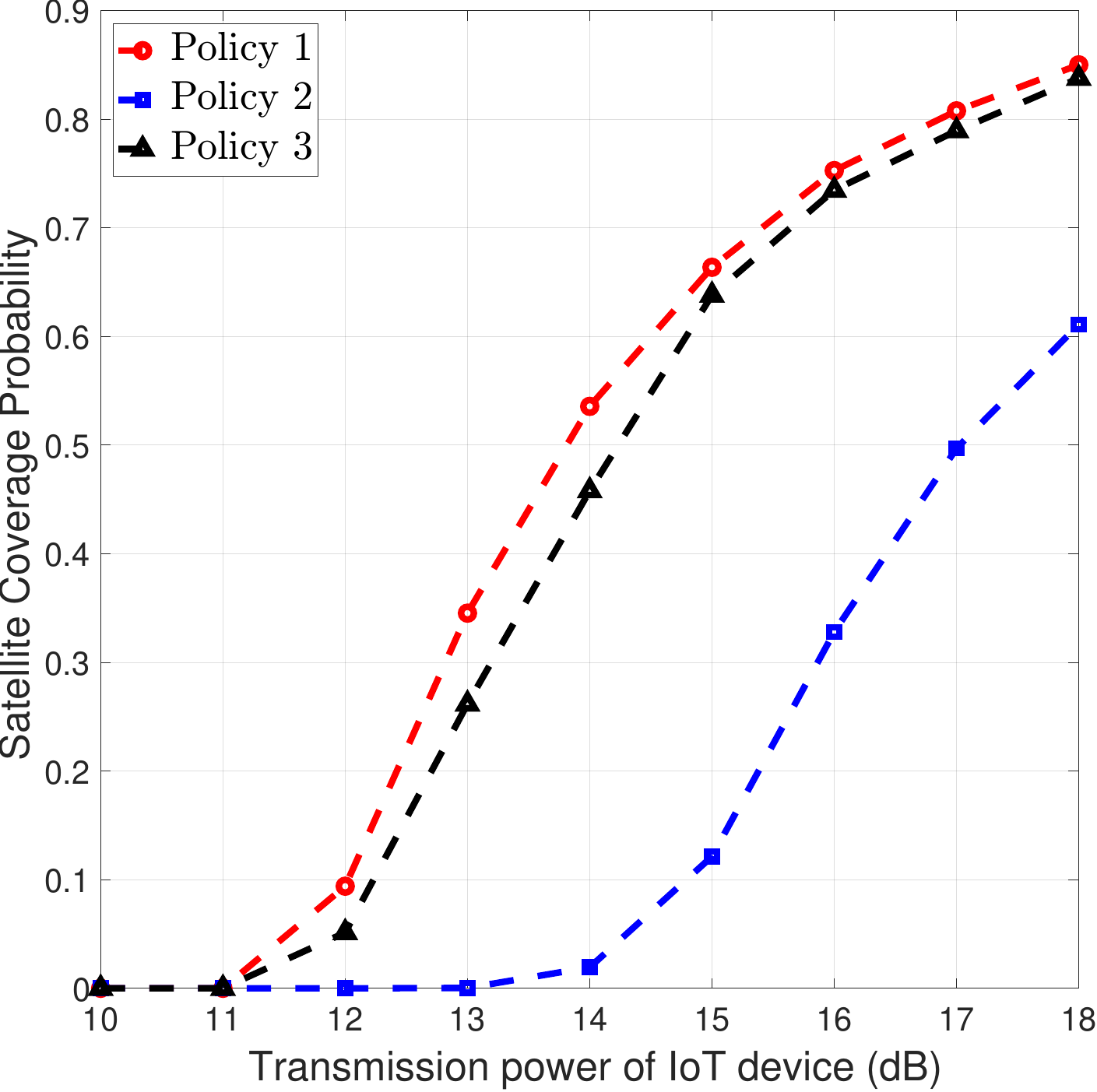}
\caption{Satellite coverage probability with different transmission policies and transmission powers.}
\label{figure8}
\end{figure}

In Fig.~\ref{figure9}, we fix the transmission power and vary the central angle to observe the impact of different transmission policies on coverage probability. The change in central angle has no impact on the position of the relay LEO satellite selected by Policy 1, so the central angle has little impact on the coverage probability for Policy 1. In contrast, the relay positions selected by Policy 2 and Policy 3 are influenced by the central angle, leading to a significant decline in their coverage performance as the central angle increases. Especially for Policy 2, when $\Theta > 20$~degrees, the chosen relay LEO satellite becomes unavailable to the IoT device, resulting in a coverage probability of $0$.

\begin{figure}[htbp]
\centering
\includegraphics[width=0.7\linewidth]{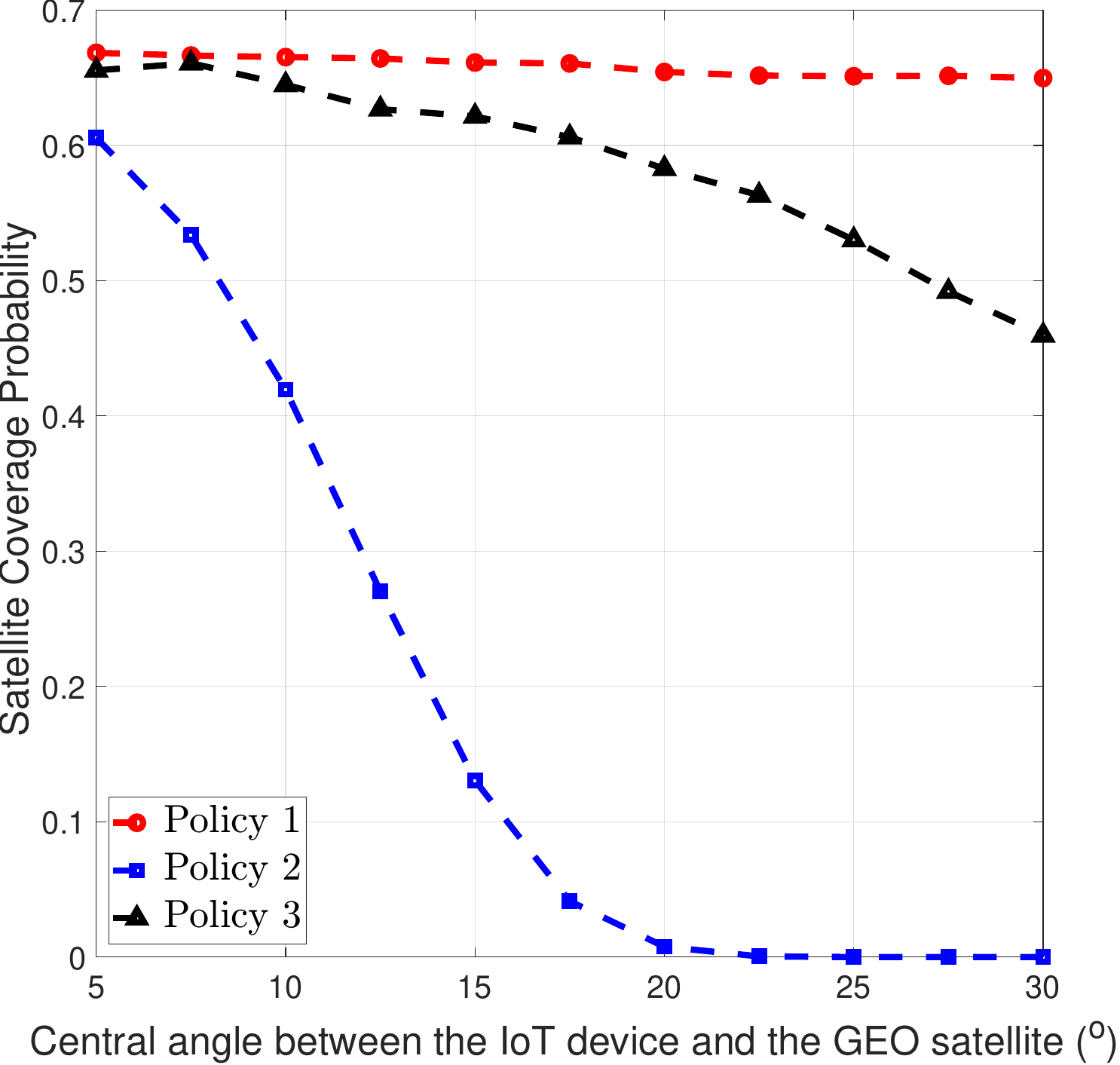}
\caption{Satellite coverage probability with different transmission policies and central angles.}
\label{figure9}
\end{figure}

\subsection{Network Architecture}
Here, we compare the advantages and disadvantages of the network architectures: $(a_1)$ IoT-LEO-GEO (proposed), $(a_2)$ IoT-GEO direct link, and $(a_3)$ IoT-LEO direct link. 

\par
\textbf{Compare $(a_1)$ and $(a_2)$:} Direct transmission from IoT devices to GEO satellites can simplify network configuration by eliminating the need for LGL synchronization. Introducing LEO satellites as relays requires additional capital and operational expenditures. Moreover, the Doppler effect is introduced with LEO satellites serving as relays, which is discussed in detail in the next subsection, is required to be overcome. However, the distance from IoT devices to GEO satellites is several dozen times greater than to LEO satellites, meaning that achieving the same coverage probability or data rate would require an increase in transmission power by several thousand times. Under the default parameters, with an IoT device transmission power of $18$~dB, nearly $100\%$ coverage probability can be achieved with LEO satellites as relays, whereas without LEO satellites, the coverage probability is approximately $0\%$. Therefore, direct communication between IoT devices and GEO satellites requires stable energy sources and powerful hardware support, which is often challenging to achieve in remote areas where IoT devices are energy-constrained.

\par
\textbf{Compare $(a_1)$ and $(a_3)$:} The path loss for a direct link between IoT devices and LEO satellites is undoubtedly smaller compared to communication with GEO satellites. Single-hop communication between IoT devices and LEO satellites improves both availability and coverage probability compared to dual-hop communication. Although the IoT-LEO direct link offers better communication performance, it is typically not a complete communication. LEO satellites have limited computational and storage capabilities, and thus they are often not the final destination for the data. Consequently, after receiving data, LEO satellites need to relay it back to ground gateways or transmit it through other satellites before it reaches the ground.

\subsection{Doppler Effect}
The rapid relative motion between satellites necessitates consideration of the Doppler shift's impact. Satellite communications typically operate at lower transmission rates, which simplifies the process of bit synchronization \cite{kaddoum2012design}. Consequently, bit synchronization adjustments can be used to mitigate the Doppler effect in satellite communications. Additionally, because bandwidth constraints are less critical in satellite systems, broadening the range of receiving frequencies does not lead to significant co-frequency interference. Therefore, spectrum spreading emerges as an effective technique to counteract Doppler shift \cite{ziedan2003bit}.
}

\section{Conclusion}
In this paper, we provide analytical expressions for the end-to-end availability and coverage performance of satellite-based IoT networks. Based on these analytical expressions, the main conclusions are as follows. Lowering the altitude of the relay, and increasing the number of satellites and transmission power can enhance network performance. The central angle between IoT devices and GEO satellites has little impact on network performance when it is small. However, once the central angle exceeds a threshold, network performance deteriorates rapidly with the central angle.

\appendices
\section{Proof of Lemma~\ref{lemma1}} \label{app:lemma1}
Considering the rotational invariance of the BPP, the PDF of the distance between the IoT device and the LEO satellite should be independent of the azimuthal angle of the LEO satellite, relying solely on the central angle. Denote $\mathcal{S}(\theta)$ as the spherical cap with a central angle as $2\theta$, the CDF of the central angle is given by
\begin{equation}
\begin{split}
    & F_{\theta_{\mathrm{LEO}}}(\theta) = \mathbb{P}\left[ \theta_{\mathrm{LEO}} \leq \theta \right] \\
    & = 1 - \mathbb{P}\left[ {\mathcal{N}\left( {{\mathcal{S}(\theta)}} \right) = 0} \right] \\
    & = 1 - \left( 1 - \frac{\mathcal{A}\left( \mathcal{S}(\theta) \right)}{\mathcal{A}\left( \mathcal{S}(\pi) \right)} \right)^{N_{\mathrm{LEO}}} \\
    & = 1 - \left( 1 - \frac{ 2\pi R_{\mathrm{LEO}}^2 (1-\cos\theta)}{4\pi R_{\mathrm{LEO}}^2} \right)^{ N_{\mathrm{LEO}}} \\
    & = 1 - \left( \frac{ 1 + \cos\theta }{2} \right)^{N_{\mathrm{LEO}}},
\end{split}
\end{equation}
where $\mathcal{N}\left( \mathcal{S}(\psi) \right)$ counts the number of satellites in the spherical cap $\mathcal{S}(\psi)$, and $\mathcal{A}\left( \mathcal{S}(\psi) \right)$ denotes the area of $\mathcal{S}(\psi)$. Then, the PDF of the central angle can be derived by
\begin{equation}
\begin{split}
    & f_{\theta_{\mathrm{LEO}}}(\theta) = \frac{\mathrm{d}}{\mathrm{d}\theta} F_{\theta_{\mathrm{LEO}}}(\theta) \\
    & = \frac{N_{\mathrm{LEO}} \sin\theta}{2} \left( \frac{ 1 + \cos\theta }{2} \right)^{N_{\mathrm{LEO}}-1}.
\end{split}
\end{equation}

\par
Finally, when the distance between the LEO relay satellite and the IoT device is $l_{\mathrm{LEO}}^{\max}$, the central angle between them reaches its upper bound. Denote the upper bound as $\theta_{\mathrm{IL}}^{\max}$, which can be expressed by the cosine rule,
\begin{equation}
    \cos \theta_{\mathrm{IL}}^{\max} = \frac{R_{\mathrm{LEO}}^2 + R_{\oplus}^2 - \left(l_{\mathrm{LEO}}^{\max}\right)^2 }{2 R_{\mathrm{LEO}} R_{\oplus} }.
\end{equation}

\section{Proof of Theorem~\ref{theorem1}} \label{app:theorem1}
Because the distance distribution of LGL is not independent of the distance distribution of ILL, we denote the probability that the distance from the LEO satellite to the GEO satellite is less than $l_{\mathrm{LG}}^{\max}$ as $P^A_{\mathrm{LG}}(\theta)$, given that the central angle between the IoT device and LEO satellite is $\theta$. From the definition, the satellite availability probability can be written as,
\begin{equation}
    P^A = \mathbbm{E}_{\theta_{\mathrm{IL}}} \left[ P^A_{\mathrm{LG}} (\theta_{\mathrm{IL}}) \right] = \int_0^{ \theta_{\mathrm{IL}}^{\max}} f_{\theta_{\mathrm{IL}}} (\theta) P^A_{\mathrm{LG}}(\theta) {\mathrm{d}}\theta.
\end{equation}

\par
Therefore, the next step is to derive the analytical expression for $P^A_{\mathrm{LG}}(\theta)$ to complete the proof. As the polar angle of the LEO satellite has been fixed as $\theta$, we need to know the range of the azimuth angle $\varphi$ of the LEO satellite when the LGL is available. When $\varphi=0$, the LEO satellite is closest to the GEO satellite, and the distance between them is
\begin{sequation}
    d_1 = \sqrt{ R_{\mathrm{LEO}}^2 + R_{\mathrm{GEO}}^2 - 2 R_{\mathrm{LEO}} R_{\mathrm{GEO}} (\sin\theta \sin\Theta + \cos\theta \cos\Theta )}.
\end{sequation}
When $d_1 > l_{\mathrm{LG}}^{\max}$, $P_{\mathrm{LG}}^A = 0$. When $\varphi=\pi$, the LEO satellite is farthest to the GEO satellite, and the distance between them is
\begin{sequation}
    d_2 = \sqrt{ R_{\mathrm{LEO}}^2 + R_{\mathrm{GEO}}^2 - 2 R_{\mathrm{LEO}} R_{\mathrm{GEO}} ( \cos\theta \cos\Theta - \sin\theta \sin\Theta)}.
\end{sequation}
When $d_2 \leq l_{\mathrm{LG}}^{\max}$, $P_{\mathrm{LG}}^A = 1$. Otherwise, when $d_1 < l_{\mathrm{LG}}^{\max} < d_2$, the distance between the LEO satellite located at $\left( R_{\mathrm{LEO}}, \theta, \varphi \right)$ and the GEO satellite is
\begin{equation}
\begin{split}
    d_3 & = \bigg( R_{\mathrm{LEO}}^2 + R_{\mathrm{GEO}}^2 - 2 R_{\mathrm{LEO}} R_{\mathrm{GEO}} \\
    & \times \left(\sin\theta \sin\Theta \cos\varphi + \cos\theta \cos\Theta \right) \bigg)^{\frac{1}{2}}.
\end{split}
\end{equation}
When $d_3 = l_{\mathrm{LG}}^{\max}$, $\varphi$ reaches its critical values
\begin{equation} 
    \cos\varphi = \frac{R_{\mathrm{LEO}}^2 + R_{\mathrm{GEO}}^2 - \left( l_{\mathrm{LG}}^{\max} \right)^2 }{2 R_{\mathrm{LEO}} R_{\mathrm{GEO}} \sin\theta \sin\Theta} - \frac{\cos\theta \cos\Theta}{\sin\theta \sin\Theta}.
\end{equation}
Since the value of $\varphi$ follows a uniform distribution, therefore
\begin{sequation}
    P_{\mathrm{LG}}^A (\theta) = \frac{1}{\pi} \arccos\left( \frac{R_{\mathrm{LEO}}^2 + R_{\mathrm{GEO}}^2 - \left( l_{\mathrm{LG}}^{\max} \right)^2 }{2 R_{\mathrm{LEO}} R_{\mathrm{GEO}} \sin\theta \sin\Theta} - \frac{\cos\theta \cos\Theta}{\sin\theta \sin\Theta} \right).
\end{sequation}

\par
Finally, we present a corollary of the above result. When $P_{\mathrm{LG}}^A (\theta)$ is replaced by $1$, LGL is always available. At this point, $P^A$ degenerates into the availability probability of ILL,
\begin{equation}
\begin{split}
    P_{\mathrm{IL}}^A & = \int_0^{ \theta_{\mathrm{IL}}^{\max}} f_{\theta_{\mathrm{IL}}} (\theta) {\mathrm{d}}\theta = F_{\theta_{\mathrm{IL}}} \left(\theta_{\mathrm{IL}}^{\max}\right) \\
    & = 1 - \left( \frac{ 1 + \cos\theta_{\mathrm{IL}}^{\max} }{2} \right)^{N_{\mathrm{LEO}}}.
\end{split}
\end{equation}


\section{Proof of Lemma~\ref{lemma2}} \label{app:lemma2}
The unconditional PDF of the pointing error can be derived based on the definition:
\begin{equation}
\begin{split}
    & f_{W_{\mathrm{LG}}} (w) = \int_0^{\infty} f_{W_{\mathrm{LG}}|\theta_d} (w) f_{\theta_d}(\theta_d) {\mathrm{d}}\theta_d \\
    & = \frac{\eta_s^2 w^{\eta_s^2  - 1 } }{A_0^{\eta_s^2}} \int_0^{\infty} \cos\theta_d \frac{\theta_d}{\varsigma^2}\exp\left ( -\frac{\theta_d^2}{2\varsigma^2} \right ) \mathrm{d} \theta_d \\
    & \overset{(a)}{\approx} \frac{\eta_s^2 w^{\eta_s^2  - 1 } }{A_0^{\eta_s^2}} \left( 1 - \frac{1}{2} \int_0^{\infty} \frac{\theta_d^3}{\varsigma^2}\exp\left ( -\frac{\theta_d^2}{2\varsigma^2} \right ) \mathrm{d}\theta_d \right) \\ 
    & \overset{(b)}{=} \frac{\eta_s^2 w^{\eta_s^2  - 1 } }{A_0^{\eta_s^2}} \left( 1 - \int_0^{\infty} z\varsigma^2  \exp\left ( -z \right ) \mathrm{d}z \right) \\
    & \overset{(c)}{=} \frac{\eta_s^2 w^{\eta_s^2  - 1 } }{A_0^{\eta_s^2}} \left( 1 - \varsigma^2 \right),
\end{split}
\end{equation}
where step $(a)$ follows the second-order Taylor expansion of $\cos\theta_d\approx 1 - \frac{\theta_d^2}{2}$, since $\theta_d$ is generally a small value. Step $(b)$ is derived by the substitution of $z = {\theta_d^2}/{2 \varsigma^2}$, and step $(c)$ is obtained by taking the expectation of the exponential distribution. 

\par
Then, the unconditional CDF obtained by integrating the  unconditional PDF,
\begin{equation}
\begin{split}
    F_{W_{\mathrm{LG}}} (w) = \int_0^w \frac{\eta_s^2 z^{\eta_s^2  - 1 } }{A_0^{\eta_s^2}} \left( 1 - \varsigma^2 \right) {\mathrm{d}}z = \frac{w^{\eta_s^2}}{A_0^{\eta_s^2}} \left( 1 - \varsigma^2 \right).
\end{split}
\end{equation}

\section{Proof of Theorem~\ref{theorem2}} \label{app:theorem2}
Due to the correlation between the distances of ILL and LGL,
\begin{equation}
\begin{split}
& P^C = \mathbbm{P} \left[ \frac{\rho_{\mathrm{IL}}^r}{\sigma_{\mathrm{LEO}}^2} > \gamma_{\mathrm{IL}}, \frac{\rho_{\mathrm{LG}}^r}{\sigma_{\mathrm{GEO}}^2} > \gamma_{\mathrm{LG}} \right] \\
& \neq \mathbbm{P} \left[ \frac{\rho_{\mathrm{IL}}^r}{\sigma_{\mathrm{LEO}}^2} > \gamma_{\mathrm{IL}} \right] \times \mathbbm{P} \left[ \frac{\rho_{\mathrm{LG}}^r}{\sigma_{\mathrm{GEO}}^2} > \gamma_{\mathrm{LG}} \right].
\end{split}
\end{equation}
Considering that, given the position of relay LEO satellite $(R_{\mathrm{LEO}},\theta_{\mathrm{IL}},\varphi)$, the conditional coverage probabilities of the two links are independent of each other. Therefore, we calculate the end-to-end coverage probability by sequentially obtaining the conditional coverage probabilities and then taking the expectation over $\theta_{\mathrm{IL}}$. From the definition, the coverage probability can be expressed as,
\begin{equation}
\begin{split}
& P^C = \mathbbm{E}_{\theta_{\mathrm{IL}},\varphi} \left[ \mathbbm{P} \left[ \frac{\rho_{\mathrm{IL}}^r}{\sigma_{\mathrm{LEO}}^2} > \gamma_{\mathrm{IL}}, \frac{\rho_{\mathrm{LG}}^r}{\sigma_{\mathrm{GEO}}^2} > \gamma_{\mathrm{LG}} \, \bigg| \, \theta_{\mathrm{IL}},\varphi \right] \right] \\
& = \int_0^{ \theta_{\mathrm{IL}}^{\max}} \int_0^{ P_{\mathrm{LG}}^A(\theta) } f_{\theta_{\mathrm{IL}}}(\theta) \frac{R_{\mathrm{LEO}}^2 \sin\theta}{4 \pi R_{\mathrm{LEO}}^2} \\
& \times \mathbbm{P} \left[ \frac{\rho_{\mathrm{IL}}^r}{\sigma_{\mathrm{LEO}}^2} > \gamma_{\mathrm{IL}}, \frac{\rho_{\mathrm{LG}}^r}{\sigma_{\mathrm{GEO}}^2} > \gamma_{\mathrm{LG}} \, \bigg| \, \theta,\varphi \right] {\mathrm{d}}\varphi {\mathrm{d}}\theta \\
& = \int_0^{ \theta_{\mathrm{IL}}^{\max}} \int_0^{ P_{\mathrm{LG}}^A(\theta) } f_{\theta_{\mathrm{IL}}}(\theta) \frac{\sin\theta}{4 \pi} \mathbbm{P} \left[ \frac{\rho_{\mathrm{IL}}^r}{\sigma_{\mathrm{LEO}}^2} > \gamma_{\mathrm{IL}} \, \bigg| \, \theta \right] \\
& \times \mathbbm{P} \left[ \frac{\rho_{\mathrm{LG}}^r}{\sigma_{\mathrm{GEO}}^2} > \gamma_{\mathrm{LG}} \, \bigg| \, \theta,\varphi \right] {\mathrm{d}}\varphi {\mathrm{d}}\theta.
\end{split}
\end{equation}
where $\theta_{\mathrm{IL}}^{\max}$ and $\pi P_{\mathrm{LG}}^A(\theta)$ are the upper bound of the LEO satellite's polar angle and azimuth angle. These two upper bounds ensure the availability of the LEO satellite.

\par
Next, we explicitly represent the first conditional probability as
\begin{equation}\label{AppD-1}
\begin{split}
& \mathbbm{P} \left[ \frac{\rho_{\mathrm{IL}}^r}{\sigma_{\mathrm{LEO}}^2} > \gamma_{\mathrm{IL}} \, \bigg| \, \theta \right] \\
& = \mathbbm{P} \left[ \frac{\rho_{\mathrm{LG}} G_{\mathrm{LG}} \zeta_{\mathrm{LG}} W_{\mathrm{LG}} }{\sigma_{\mathrm{LEO}}^2} \left(\frac{4\pi}{\lambda_{\mathrm{LG}}} l_{\mathrm{LG}}\right)^2 > \gamma_{\mathrm{LG}} \, \bigg| \, l_{\mathrm{LG}}^2 = R_{\mathrm{LEO}}^2 + R_{\oplus}^2 - 2R_{\mathrm{LEO}} R_{\oplus} \cos\theta \right] \\
& = \mathbbm{P} \left[ W_{\mathrm{LG}} > \frac{\gamma_{\mathrm{IL}} \sigma_{\mathrm{LEO}}^2 }{\rho_{\mathrm{IL}}^t G_{\mathrm{IL}} \zeta_{\mathrm{IL}} } \left( \frac{4 \pi}{\lambda_{\mathrm{IL}}} \right)^2 \left( R_{\mathrm{LEO}}^2 + R_{\oplus}^2 - 2R_{\mathrm{LEO}} R_{\oplus} \cos\theta \right) \right] \\
& = 1 - F_{W_{\mathrm{LG}}} \left( \frac{\gamma_{\mathrm{IL}} \sigma_{\mathrm{LEO}}^2 }{\rho_{\mathrm{IL}}^t G_{\mathrm{IL}} \zeta_{\mathrm{IL}} } \left( \frac{4 \pi}{\lambda_{\mathrm{IL}}} \right)^2 \left( R_{\mathrm{LEO}}^2 + R_{\oplus}^2 - 2R_{\mathrm{LEO}} R_{\oplus} \cos\theta \right) \right).
\end{split}
\end{equation}
As for the derivation of the second conditional probability, given the position of LEO satellite is $(R_{\mathrm{LEO}},\theta,\varphi)$, the distance between the LEO satellite and GEO satellite is $d \left( R_{\mathrm{GEO}},\Theta,0; R_{\mathrm{LEO}},\theta,\varphi \right)$. The specific explicit expression of the operator $d \left( \cdot,\cdot,\cdot; \cdot,\cdot,\cdot \right)$ is provided in (\ref{Euclidean}).  The remaining parts follow a similar procedure and, therefore omitted here. 

\bibliographystyle{IEEEtran}

{\color{black}
\bibliography{references}
}

\end{document}